\documentclass[twocolumn,showpacs,preprintnumbers,amsmath,amssymb]{revtex4}
\usepackage{graphicx}% Include figure files
\usepackage{dcolumn}% Align table columns on decimal point
\usepackage{bm}% bold math

\begin{document}

%\preprint{APS/123-QED}

\title{Growth mechanisms of GaN on the O-terminated ZnO(000\={1}) surfaces}% Force line breaks with \\

\author{Katsutoshi Fujiwara$^1$, Akira Ishii$^{1,2}$, Toshikazu Ebisuzaki$^3$, Atsushi Kobayashi$^4$, Yuji Kawaguchi$^4$, Jitsuo Ohta$^{4,5}$, Hiroshi Fujioka$^{4,5}$}
% \email{fujiwara@kjaro.damp.tottori-u.ac.jp}
% \altaffiliation{}%Lines break automatically or can be forced with \\
\affiliation{
1. Department of Applied Mathematics and Physics, Tottori University, Koyama, Tottori, 680-8552, Japan \\
2. National Institute of Advanced Industrial Science and Technology (AIST), Japan \\
3. Computational Astrophysics Laboratory, RIKEN, 2-1 Hirosaki, Wako, Saitama 351-0198, Japan \\
4. Institute of Industrial Science, The University of Tokyo, 4-6-1 Komaba, Meguro-ku, Tokyo 153-8505, Japan \\
5. Kanagawa Academy of Science and Technology (KAST), Japan
}

\date{\today}% It is always \today, today,

\begin{abstract}
We have investigated the stability of the 1ML-GaN on the O-polarity ZnO(000\={1}) interface structure using the first-principles calculation. We have found in our calculated results that the most stable structure for the 1ML-GaN on the O-polarity ZnO(000\={1}) interface has the N-polarity. However, we have found that the results of the adatom dynamics on the O-terminated ZnO(000\={1}) surface shows the Ga-polarity. We find that the key to change the polarity of GaN crystal grown on the O-terminated ZnO(000\={1}) surface is the growth temperature. We have suggested that the optimized initial growth condition for the growth of the Ga-polarity GaN crystal on the O-terminated ZnO(000\={1}) surface is under the suitable low temperature and the stoichiometric growth condition. Experimental observations show that GaN grown on ZnO(000\={1}) by PLD at substrate temperatures below 300$^\circ\mathrm{C}$ has Ga-polarity, which is quite consistent with the theoretical calculations.
\end{abstract}

%Use showkeys class option if keyword
\maketitle

%%%%%%%%%%%%%%%%  Section  %%%%%%%%%%%%%%%%%%%%
\section{\label{sec:level1}Introduction}
GaN has attracted much attention as a material for short-wavelength optical devices such as blue light-emitting diodes (LED) and laser diodes (LD) since it has a direct band gap of 3.4 eV.\cite{1,2} Although, GaN films are usually grown on sapphire substrates, it is well known that these GaN films suffer from formation of a high density of structural defects due to the large lattice mismatch between GaN and sapphire. ZnO has been regarded as one of the most promising substrates because ZnO and GaN perfectly share the same crystalline symmetries and the lattice mismatches between them are as small as 1.9\% and 0.4\% 
for the a-axis and c-axis, respectively. In fact, Kobayashi et al.\cite{3} have recently shown that high quality GaN grows in the layer-by-layer mode on ZnO(000\={1}) even at room temperature (RT) by the use of pulsed laser deposition (PLD). They have also shown that the surface of GaN grown on ZnO(000\={1}) has straight steps with a height of one unit cell and large atomically flat terraces. These striking features can be attributed to the small lattice mismatch between GaN and ZnO. In spite of this success in heteroepitaxial growth, the initial stage of the film growth of GaN on ZnO(000\={1}) has not been understood very well. Namkoong et al.\cite{4} have recently reported that MBE growth of GaN on atomically flat O-terminated ZnO(000\={1}) substrates at a substrate temperature of 600$^\circ\mathrm{C}$ leads to formation of mixed polarity GaN. However, existence of large atomically flat terraces that Kobayashi et al.\cite{3} have reported for RT grown GaN implies possibility of growth of Ga-polarity GaN. Further study is necessary to clarify the initial stage of film growth and to understand mechanisms that determines the polarity of GaN grown on ZnO(000\={1}) substrates.

First-principle calculation has been proved to be a powerful tool for investigation of growth mechanisms of group III nitrides. For example, the first-principle calculations have been successfully applied to investigate mechanisms for the Ga-terminated GaN(0001) homoepitaxial growth.\cite{5,6} Polarity change of GaN by the use of the Al-terminated sapphire(0001) has also been well explained by the first-principles calculation.\cite{7} 
In this study, we discuss mechanisms of the polarity determination of GaN on the O-terminated Zn(000\={1}) surface with the theoretical first-principles calculations of the total energy for the 1ML-GaN on the O-terminated ZnO(000\={1}) interfacial structure. 

%%%%%%%%%%%%%%%%  Section  %%%%%%%%%%%%%%%%%%%%
\section{Computational method}
Our first-principles density-functional total-energy calculations are based on the density-functional theory(DFT),\cite{8,9} using the Vienna ab initio simulation package (VASP).\cite{10} We used the local-density approximation(LDA)\cite{11,12} for the exchange correlation and projector augmented wave (PAW)\cite{13,14} potentials. The Zn and Ga 3d electrons were treated as part of the valence band. The cutoff energy for the plane-wave basis was 400 eV. The twin boundaries are modeled with supercells. We employed supercells containing eight atomic layers of ZnO and two layers of GaN, a vacuum region 12 \AA ~thick. The cation-terminated bottom layer of the slab was passivated with fractionally charged hydrogen atoms. Two layers on this side of the slab were kept fixed to simulate the constrains on the few outmost layers coming from the underlying semi-infinite bulk, whereas all other atoms were allowed to relax until the Hellmann-Feynman forces became smaller than 0.05 eV/\AA. We employed sets of special k points equivalent to 4 kpoints within the surface Brillouin zone (BZ).

%%%%%%%%%%%%%%%%  Section  %%%%%%%%%%%%%%%%%%%%
\section{Results and discussion}
\begin{figure}
\includegraphics[width=8cm]{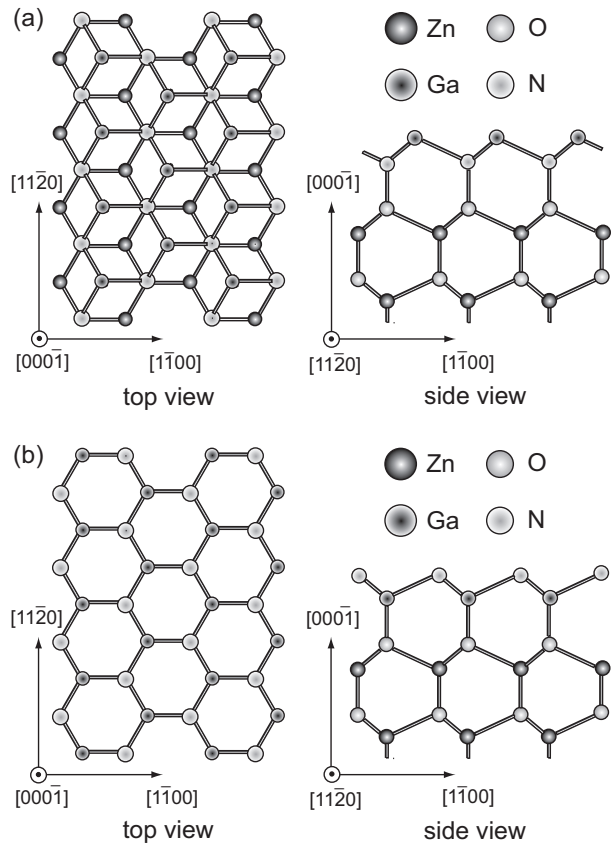}% Here is how to import EPS art
\caption{\label{fig:1ML_mono} Atomic structure (top and side view) for the 1ML-GaN on the O-terminated ZnO(000\={1}) surface. (a) is the Ga-polarity zincblende interfacial structure, and (b) is the N-polarity wurtzite interfacial structure.}
\end{figure}

We calculated the total energy for the 1ML-GaN on the O-terminated ZnO(000\={1}) surface. We calculated four structural models that the polarity has the Ga-polarity and N-polarity, and the interfacial structure has the Zincblend type and Wurtzite type structure each other shown in Fig. ~\ref{fig:1ML_mono}. We employed (1$\times$1) periodicity, fixing the 1ML-GaN laterally and height to relax calculate the total energy. We were shown that the N-polarity structure was more stable than the Ga-polarity structure in Table~\ref{tab:table1}. The most stable for the 1ML-GaN on the O-terminated ZnO(000\={1}) interfacial structure was N-polarity Wurtzite type interfacial structure shown in Figure ~\ref{fig:1ML_mono}(b). The relative energy of the Ga-polarity Zincblend type interfacial structure shown in Fig.~\ref{fig:1ML_mono}(a) that was suggested from atomic diffusion was 0.90eV. Therefore, it was shown that the GaN growth layer on the O-terminated ZnO(000\={1}) surface has the N-polarity Wurtzite interfacial structure. It was shown that both the Ga-polarity and N-polarity GaN possibilities of grown on the O-terminated ZnO(000\={1}) surface using the first principle calculation. We discuss the cause into which the polarity changes by paying attention to the migration paths and the migration barrier energies. We are given the structure of the O-terminated ZnO(000\={1}) surface, we focus our attention on the relative stability of the four high-symmetry sites, namely, the original wurtzite and zincblende lattice site (L), the bridge site (B), the hollow and three neighbors site (H3), and the top of the second layer and four neighbors site (T4).
%%%%%%%%%%%%%%%%  Fig  %%%%%%%%%%%%%%%%%%%%
%%%%%%%%%%%%%%%%  Fig  %%%%%%%%%%%%%%%%%%%%
%%%%%%%%%%%%%%%%  Tables  %%%%%%%%%%%%%%%%%%%%
\begin{table}
\caption{\label{tab:table1}Total energy valuses of 1ML-GaN on the O-terminated ZnO(000\={1}) interfacial structure.The energy values for the stable structure is set at zero. The unit of the energy is eV.}
\begin{ruledtabular}
\begin{tabular}{cc}
GaAs/ZnO(000\={1}) structure & Relative energy (eV/1$\times$1) \\
\hline
Ga-polarity Wurtzite & 0.93 \\
Ga-polarity Zincblende & 0.90 \\
N-polarity Wurtzite & 0.00 \\
N-polarity Zincblende & 0.08 \\
\end{tabular}
\end{ruledtabular}
\end{table}
%%%%%%%%%%%%%%%%  Tables  %%%%%%%%%%%%%%%%%%%%

For Ga adatoms the most stable adsorption site on the O-terminated ZnO(000\={1}) surface is the H3 site. The energetically lowest transition path is the B site and the T4 site. The T4 site is significantly higher in energy of 0.49 eV. Therefore, to hop from one T4 site to the next, the Ga adatom diffuses along the B site over the T4 site which is the transition site leading to the migration barrier energy of 0.49 eV. The L site is unstable for adsorption of Ga adatom: the relative energy is 0.96 eV. The origin of the energy difference comes from the fact that an isolated Ga atom has three bonds. At the H3 site, the Ga adatom uses all three bonds with three O atoms of the topmost layer. However, at the L site, the Ga adatom uses only one bond for adsorbation and the remaining two bonds are dangling-bonds. Thus, the crystal growth will be very difficult after constructing such a structure.

For N adatoms the most stable adsorption site on the O-terminated ZnO(000\={1}) surface is L site. The energetically lowest transition path is the B site. The B site is significantly higher in energy of 0.15 eV. Therefore, to hop from one L site to the next, the N adatom diffuses along the B site which is the transition site leading to the migration barrier energy of 0.15 eV. The origin of the energy difference comes from the fact that an isolated N atom has three bonds. At the H3 and T4 sites, the N adatom uses all three bonds with three O atoms of the topmost layer. However, the N-O bond length is shorter than the Zn-O bond length. 

%%%%%%%%%%%%%%%%  Tables  %%%%%%%%%%%%%%%%%%%%
\begin{table}
\caption{\label{tab:table2}The migration barrier energy values of the N on the Ga-terminated GaN(0001) surface and the O-terminated ZnO(000\={1}) surface. The unit of the energy is eV.}
\begin{ruledtabular}
\begin{tabular}{cc}
surface & Migration barrier energy (eV) \\
\hline
 & Ga adatom ~ N adatom \\
\hline
GaN(0001)\footnote{Ref.\cite{5}.} & 0.4~ ~~~~~~~~ 1.4 ~\\
ZnO(000\={1})\footnote{Ref.\cite{4}.} &~0.49 ~~~~~~~~0.15 \\
\end{tabular}
\end{ruledtabular}
\end{table}
%%%%%%%%%%%%%%%%  Tables  %%%%%%%%%%%%%%%%%%%%

Thus, the N adatom is bind to one O atom of the topmost layer. According to the calculation, an isolated N adatom is the most stable at the L site, and an isolated Ga adatom is unstable at the L site.
 Table~\ref{tab:table2} show the migration barrier energies of the Ga and N adatoms were compared with them on the Ga-terminated GaN(0001) surface that have been investigated theoretically by Zywietz et al.5 It was shown that the migration barrier energy of the Ga adatom was similar to on the Ga-terminated GaN(0001) surface by 0.4eV. However, It was shown that the migration barrier energy of the N adatom was very smaller than on the Ga-terminated GaN(0001) surface by 1.4eV. Thus, the higher possibility of the initial optimized growth condition for the heteroepitaxy growth of GaN on the O-terminated ZnO(000\={1}) surface is the Ga-polarity GaN crystal grown under the suitable low temperature condition.

It was known to obtain the GaN crystal of the high quality when the GaN was grown under the Ga-rich growth condition. However, N adatom on the O-terminated ZnO(000\={1}) surface diffuses compared with on the Ga-terminated GaN(0001) surface. Therefore, we have suggested that the adsorbed atomic specie of the first layer on the O-terminated ZnO(000\={1}) surface is N-polarity because Ga adatom gives priority under the Ga-rich condition. Not Ga-rich growth condition but stoichiometric growth condition of the growth ratio is better. In general, when the GaN was grown on the Ga-polarity, it was known to obtain the GaN crystal of the high quality. As for temperature of the initial growth condition, we were suggested that the stoichiometric growth condition and the suitable low temperature condition be necessary compared with the typical growth temperature from those calculated results.

\begin{figure}
\includegraphics[width=8cm]{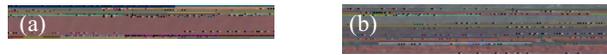}% Here is how to import EPS art
\caption{\label{fig:AFM}A RHEED pattern for GaN grown at 300$^\circ\mathrm{C}$ on a ZnO(0001) stepped and tellaced substrate. The $\times$2 reconstructed surface is indicative of a Ga-polarity surface. The incidence of the electron beam is parallel to the GaN[11\={2}0] direction.}
\end{figure}

\begin{figure}[hbtp]
\includegraphics[width=6cm]{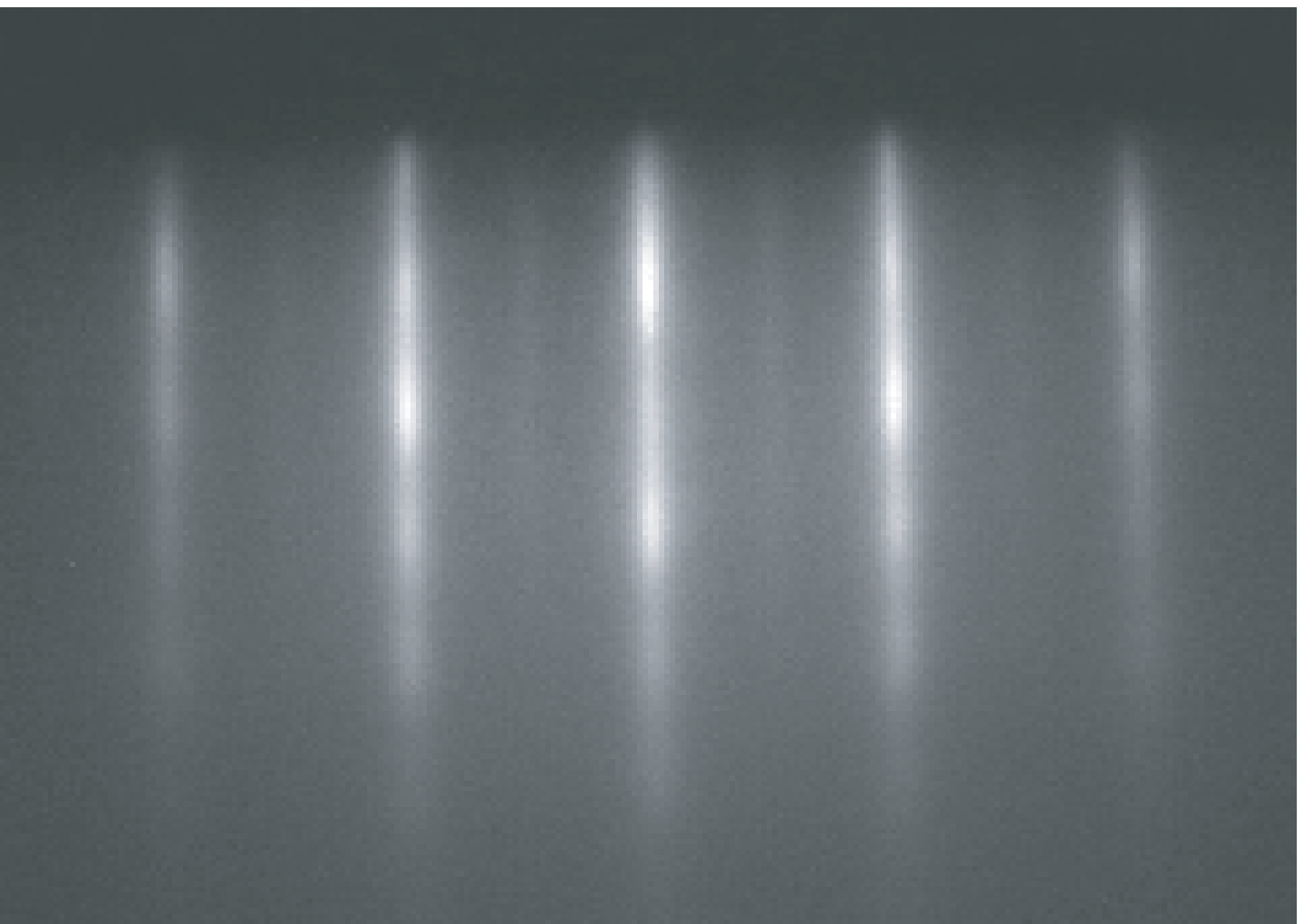}% Here is how to import EPS art
\caption{\label{fig:RHEED}1$\times$1 $\mu$m$^2$ AFM images of surfaces of GaN grown at RT (a) before and (b) after naOH etching.}
\end{figure}

We have experimentally investigated the polarity of GaN grown at low substrate temperatures under stoichiometric conditions by the use of PLD. Figure \ref{fig:AFM} shows a RHEED pattern for GaN grown at 300$^\circ\mathrm{C}$ on a ZnO stepped and terraced substrate. GaN grown at this temperature shows a sharp streaky RHEED pattern, which indicates that it has high crystallinity and a flat surface. In addition, we have also observed RHEED oscillation during the growth at 300$^\circ\mathrm{C}$, which implies that the growth proceeds in the layer-by-layer mode. It should be noted that RHEED image shown in Fig.~\ref{fig:AFM} has a $\times$2 reconstructed pattern, which is indicative of the Ga-polarity surface. Figures \ref{fig:RHEED} (a) and (b) show 1$\times$1 $\mu$m$^2$ AFM images of the surfaces of GaN grown at RT before and after NaOH etching, respectively. A stepped and terraced structure can be seen in Fig.~\ref{fig:RHEED} (a), which is consistent with the layer-by-layer growth mode. One can see that the surface of GaN remains atomically flat even after the NaOH etching, which makes a concrete evidence for growth of the Ga-polarity GaN. The facts that GaN grown at substrate temperatures below 300$^\circ\mathrm{C}$ under stoichiometric conditions has Ga-polarity and GaN grown at 600$^\circ\mathrm{C}$ has mixed polarity as reported by Namkoong et al.\cite{4} are quite consistent with our theoretical calculations.

%%%%%%%%%%%%%%%%  Fig  %%%%%%%%%%%%%%%%%%%%

%%%%%%%%%%%%%%%%  Fig  %%%%%%%%%%%%%%%%%%%%

%%%%%%%%%%%%%%%%  Section  %%%%%%%%%%%%%%%%%%%%
\section{Summary}
In summary, we have investigated the stability of the 1ML-GaN on the O-polarity ZnO(000\={1}) interface structure using the first-principles calculation. We have found that the calculated results of the most stable of the 1ML-GaN on the O-polarity ZnO(000\={1}) interface has the N-polarity. However, we have found that the calculated results of the adatom diffusion on the O-terminated ZnO(000\={1}) surface has the Ga-polarity. Therefore, the change in the polarity of GaN crystal grown on the O-terminated ZnO(000\={1}) surface was important of the change of growth temperature. We have suggested that the optimized initial growth condition for the growth of the Ga-polarity GaN crystal on the O-terminated ZnO(000\={1}) surface is under the stoichiometric growth rate and the suitable low temperature condition. Experimental observations show that GaN grown on ZnO(000\={1}) by PLD at substrate temperatures below 300$^\circ\mathrm{C}$ has Ga-polarity, which is quite consistent with the theoretical calculations.

%%%%%%%%%%%%%%%%  acknowledgments  %%%%%%%%%%%%%%%%%%%%
\begin{acknowledgments}
The computational calculations are performed at High-Performance Super Computer Center of RIKEN Institute. One of the authors (A.I.) is also grateful for the Grand-in-Aide for Scientific research of the Japan Society for the Promotion of Science for travel fee support for joint research.
\end{acknowledgments}

\newpage %Just because of unusual number of tables stacked at end
\bibliography{apssamp}% Produces the bibliography via BibTeX.

\end{document}